\documentclass[aps,prl,reprint,groupedaddress]{revtex4-1}
\usepackage{amsmath}
\usepackage{amssymb}
\usepackage{graphicx}
\usepackage{bm}

\begin{document}

%\preprint{}

\title{Localized and intense energy conversion in the diffusion region
  of asymmetric magnetic reconnection}


\author{M. Swisdak$^1$} \author{J. F. Drake$^1$}
\author{L. Price$^1$} \author{J. L. Burch$^2$}
\author{P. A. Cassak$^3$} \author{T.-D. Phan$^4$} \affiliation{$1.$
  University of Maryland, College Park, MD 20742, USA\\ $2.$ Southwest
  Research Institute, San Antonio, TX, USA\\ $3.$ West Virginia
  University, Morgantown, WV, USA\\ $4.$ Space Sciences Laboratory,
  University of California, Berkeley, California, USA}
\email[]{swisdak@umd.edu}
%\thanks{}
%\affiliation{}

\date{\today}

\begin{abstract}
We analyze a high-resolution simulation of magnetopause reconnection
observed by the Magnetospheric Multiscale (MMS) mission and explain
the occurrence of strongly localized dissipation with an amplitude
more than an order of magnitude larger than expected.  Unlike
symmetric reconnection, wherein reconnection of the ambient reversed
magnetic field drives the dissipation, we find the annihilation of the
self-generated, out-of-plane (Hall) magnetic field plays the dominant
role.  Electrons flow along the magnetosheath separatrices, converge in the
diffusion region, and jet past the X-point into the magnetosphere.
The resulting accumulation of negative charge generates intense
parallel electric fields that eject electrons along the magnetospheric
separatrices and produce field-aligned beams.  Many of these features
match MMS observations.
\end{abstract}

%\keywords{}

\maketitle

\section{Introduction}\label{intro}

Magnetic reconnection transfers energy from the magnetic field to the
surrounding plasma.  Oppositely directed components of the field
undergo topological reordering at X-points, which form within electron
diffusion regions where the magnetic field is no longer frozen into
any component of the plasma. In the simplest (symmetric) case the two
plasmas upstream from an X-point differ only in the orientation of the
embedded magnetic field and the electron diffusion region is elongated
along the outflow direction but otherwise relatively unstructured
\citep{shay07a}.

However reconnection also occurs in asymmetric configurations in which
the abutting plasmas differ in density, temperature and magnetic field
strength \citep{sonnerup81a,cassak07a}, e.g., Earth's magnetopause,
the boundary separating the hot and tenuous plasma of the
magnetosphere from the magnetosheath and its shocked plasma of solar
wind origin.  The structure of asymmetric reconnection, and
magnetopause reconnection in particular, has remained unclear, with
some arguing that a localized electron diffusion region does not
develop \citep{pritchett09a,mozer09a}, even in the special case where
the opposing fields are anti-parallel.

Recent observations by the Magnetospheric Multiscale (MMS) mission
have triggered new interest in the structure of the electron diffusion
region during asymmetric reconnection. Crescent-shaped electron
velocity-space distributions that had been predicted by simulations
\citep{hesse14a} were seen \citep{burch16a,chen16a} and the role of
the large normal electric field, which points sunward across the
magnetopause and balances the ambient ion pressure gradient, in
driving the crescents to higher energies was established
\citep{bessho16a,chen16a,egedal16a,shay16a}. In contradiction to
earlier models and observations, the data suggest that energy
conversion within the electron diffusion region is associated with
oblique whistler-like disturbances featuring intense parallel electric
fields and oscillations in $\mathbf{J}\boldsymbol{\cdot}\mathbf{E}$ of
unknown origin. The local dissipation rate is nearly two orders of
magnitude greater than expected \citep{burch18a}.

Here we use a high-resolution particle-in-cell simulation to explore
the structure of asymmetric reconnection in a system with initial
conditions based on MMS observations \citep{burch16a,burch18a}. We
demonstrate that a jet of electrons streaming towards the
magnetosphere and across the X-point produces a standing structure
with non-uniform but intense energy conversion (such structures can
also be seen, although are not explained, in \cite{cassak17a}).
Unlike in symmetric reconnection, the energy transfer arises from the
annihilation of the Hall (out-of-plane) component of the magnetic
field.  The most significant annihilation does not occur at the
X-point but instead is shifted towards the magnetosphere and the fluid
stagnation point.  The electrons are ejected from the diffusion region
by intense parallel electric fields and not, as in the symmetric case,
by the reconnected magnetic field.  The simulation reproduces the key
features of the observations.

\section{Simulations}\label{sims}

We perform the simulations with the particle-in-cell code {\tt p3d}
\citep{zeiler02a}.  In its normalization a reference magnetic field
strength $B_0$ and density $n_0$ define the velocity unit
$v_{A0}=B_0/\sqrt{4\pi m_in_0}$.  Times are normalized to the
inverse ion cyclotron frequency $\Omega_{i0}^{-1}=m_ic/eB_0$, lengths
to the ion inertial length $d_{i0} =c/\omega_{pi0}$ (where
$\omega_{pi0} = \sqrt{4\pi n_0 e^2/m_i}$ is the ion plasma frequency),
electric fields to $v_{A0}B_0/c$, and temperatures to $m_iv_{A0}^2$.
In the system considered here $B_0$ and $n_0$ correspond to their
asymptotic magnetosheath values: $B_0 = 23\text{ nT}$ and $n_0 =
11.3\text{ cm}^{-3}$.

The initial conditions closely mimic those observed during the
diffusion region encounter described in \citet{burch16a}.  We employ
an $LMN$ coordinate system in which the reconnecting field parallels
the $L$ axis (roughly north-south), the $M$ axis runs roughly
east-west, with dawnward positive, and the $N$ axis points radially
away from Earth and completes the right-handed triad.  The
reconnecting component of the field $B_L$ and the ion and electron
temperatures, $T_i$ and $T_e$, vary as functions of $N$ with
hyperbolic tangent profiles of width 1. The asymptotic values of $n$,
$B_L$, $T_i$, and $T_e$ in code units are 1.0, 1.0, 1.37, and 0.12 in
the magnetosheath and 0.06, 1.70, 7.73, and 1.28 in the magnetosphere.
Pressure balance determines the initial density profile.  The guide
field $B_M=0.099$ is much smaller than $B_L$ (i.e., the reconnection
is nearly anti-parallel) and initially uniform.  While not an exact
kinetic equilibrium, the unperturbed configuration is in force balance
and would not undergo significant evolution during the timescales
considered here.  We impose a small initial perturbation in order to
trigger reconnection at a single and specific point.

The ion-to-electron mass ratio is chosen to be $100$, which is
sufficient to separate the electron and ion scales (the electron
inertial length $d_{e0} = 0.1d_{i0}$).  The normalized speed of light
is $c=15$ so that $\omega_{pe}/\Omega_{ce}=1.5$ in the asymptotic
magnetosheath and $\approx 0.2$ in the asymptotic magnetosphere; the
observed ratios are larger, $\approx 46$ and $7$, and as a consequence
the simulation's Debye length is larger than in the real system.
However, since the development of reconnection does not appreciably
depend on physical effects at the Debye scale the expected impact is
minimal.  The spatial grid has resolution $\Delta = 0.01$ in
normalized units while the Debye length in the simulation's
magnetosheath, $\approx 0.03$, is the smallest physical scale.  To
ameliorate numerical noise, particularly in the low-density
magnetosphere, each grid cell initially contains 3000 weighted
macroparticles, substantially more than typical PIC simulations.

%The
%amplitude of statistical fluctuations scales inversely with the square
%root of the number of macroparticles.

The computational domain of the principal simulation discussed here
has dimensions $(L_L,L_N) = (40.96,20.48)$ with periodic boundary
conditions used in all directions.  While particles can move in the
$M$ direction, variations in physical quantities are not permitted,
$\partial/\partial M = 0$.  This simplification greatly eases the
computational burden while still allowing reconnection to proceed.  We
also compare some of our results with those from a fully
three-dimensional simulation of the same event previously described in
\cite{price16a}.

\section{Results} 

We first discuss the case of symmetric reconnection in order to
provide background for the significantly larger energy conversion
rates seen in the observations and the simulations.  Direct
manipulation of Maxwell's equations gives Poynting's theorem:
\begin{equation}
  \frac{\partial U}{\partial t} +
\frac{c}{4\pi}\boldsymbol{\nabla\cdot}(\mathbf{E}\boldsymbol{\times}\mathbf{B})
= -\mathbf{J}\boldsymbol{\cdot}\mathbf{E},
\label{eqn:energy}
\end{equation}
where $U = (E^2 + B^2)/8\pi$ is the electromagnetic energy density,
the second term defines the Poynting flux $\mathbf{S} =
c\,(\mathbf{E}\boldsymbol{\times}\mathbf{B})/4\pi$, and
$\mathbf{J}\boldsymbol{\cdot}\mathbf{E}$ quantifies the rate of energy
transfer, being positive when directed from the fields to the
particles.  During symmetric reconnection the $M$ component of the
electric field drives an elongated (in the $L$ direction) layer of
electron current density $J_M$.  The product $J_ME_M$ matches the
divergence of the incoming Poynting flux associated with the
reconnecting field ($\sim E_MB_L)$.  However this scaling cannot
explain MMS measurements of magnetopause reconnection.  For reasonable
parameters -- current density $J_M\sim 1\text{
}\mu\text{A/}\text{m}^2$ and reconnection electric field $E_M\sim
0.2\text{ mV/m}$ -- the resulting $J_ME_M \sim 0.2\text{
  nW/}\text{m}^3$ greatly underestimates the observations, which can
exceed $10\text{ nW/}\text{m}^3$
\citep{burch16a,ergun16a,eriksson16a}.  Hence, the energy conversion
processes that dominate during symmetric reconnection likely do not
play a significant role at the magnetopause.

\begin{figure*}
\includegraphics[width=\textwidth]{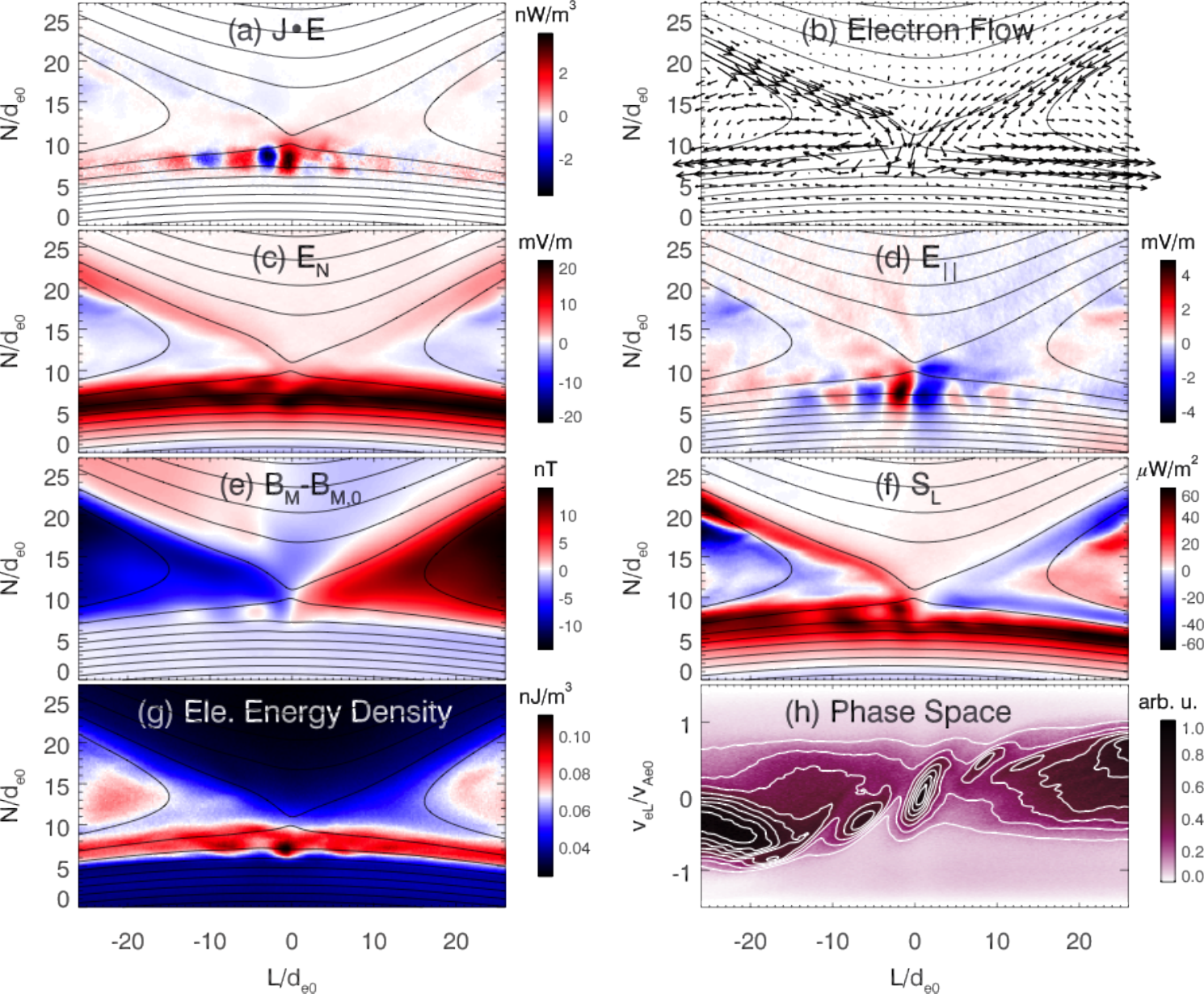}
\caption{\label{jdotefig}Simulation results from a region centered on
  the X-point. (a) The $\mathbf{J}\boldsymbol{\cdot}\mathbf{E}$ term
  from Poynting's theorem; (b) In-plane electron flow field; (c)
  $E_N$, the normal component of the electric field; (d)
  $E_{\parallel}$, the component of the electric field parallel to the
  magnetic field; (e) $B_M-B_{M,0}$, the change in the out-of-plane
  component of the magnetic field from its (spatially constant)
  initial value; (f) $S_L$, the horizontal component of the Poynting
  flux; (g) Electron energy density; (h) $v_{L}-L$ electron phase
  space overlaid with contours.  Magnetic field lines have been
  overplotted in panels (a)-(g).  Distances are normalized to the
  electron inertial length $d_{e0} = d_{i0}\sqrt{m_e/m_i} =
  0.1d_{i0}$.}
\end{figure*}

We now show that high-resolution simulations of asymmetric
reconnection can produce the large rates of energy conversion and then
discuss the underlying physical mechanism.  Figure \ref{jdotefig}
gives an overview of the simulation at $t=32 \Omega_{ci}^{-1}$, a time
of steady-state reconnection.  The magnetosheath lies at the top of
each panel and the magnetosphere at the bottom (equivalently,
earthward is down) while the horizontal axis roughly points north-south
because the MMS encounter occurred near the equatorial plane.  The
reconnecting component of the field is in the $+L$ direction in the
magnetosphere and the $-L$ direction in the magnetosheath;
representative field lines are overplotted in panels (a)-(g).  The
color bar labels have been converted from the simulation's
normalization to MKS units.

Panel (a) shows the structure of $\mathbf{J}
\boldsymbol{\cdot}\mathbf{E}$.  Strikingly, regions of positive (red)
and negative (blue) sign co-exist, the latter representing a transfer
of energy from the plasma to the fields, the opposite of the usual
behavior during reconnection.  The rate of energy conversion in the
electron frame \citep{zenitani11a},
$\mathbf{J}\boldsymbol{\cdot}\mathbf{E}^{\prime} =
\mathbf{J}\boldsymbol{\cdot}(\mathbf{E} +
\mathbf{v_e}\boldsymbol{\times}\mathbf{B}/c)$, is essentially
identical because the electron contribution dominates the current
density \citep{cassak17a}.  Although the X-point slowly drifts in the
$L$ direction during the simulation, the structure of
$\mathbf{J}\boldsymbol{\cdot}\mathbf{E}$ undergoes minimal temporal
evolution and remains stationary in the X-point frame.

Unlike in symmetric reconnection, the primary contribution to the
dissipation comes from $J_NE_N$.  High-density magnetosheath plasma
flows in the $N$ direction across the X-point and into the low-density
magnetosphere.  (As is generally the case in asymmetric reconnection,
the magnetic X-point and the flow stagnation point do not coincide
\citep{cassak07a}.)  The $v_NB_L$ Lorentz force redirects the motion
of both the electrons and the ions, but the larger masses of the
latter let them penetrate farther into the magnetosphere.  The
resulting charge imbalance produces a large $E_N$ that retards the ion
motion and balances the ion pressure gradient \citep{pritchett08a}
while also accelerating electrons towards the magnetosphere. Panel (b)
shows the electron flow and, in particular, the $N$-directed flow
across the X-point. 
%Panel (c) shows the resulting charge imbalance,
%with blue corresponding to an excess of electrons.  Although the
%magnitude of the net charge density, $\approx 10\%$ of the charge
%density of each individual species, is unrealistically large because
%of the simulation's unrealistic speed of light and Debye length, its
%structure should be representative.  The electric field amplitudes,
%however, will be typical since they are controlled by the dynamics of
%the plasma at spatial scales greater than the Debye length.  
Panel (c) shows the strong positive magnetospheric $E_N$ that results
from the separation of positive and negative charge.  This separation
occurs along the entire magnetospheric separatrix due to the different
spatial scales associated with electrons and ions.

The interaction of the current due to the electron flow with $E_N$
produces the large region of positive
$\mathbf{J}\boldsymbol{\cdot}\mathbf{E}$ near the center of panel (a).
Since $J_N \sim J_M$ in the simulation, it is the large value of $E_N
\sim 20\text{ mV/m} \gg E_M$ that makes the most significant
contribution to the energy transfer rate and causes the local
dissipation rate to greatly exceed expected values. We return to the
energy source of this intense dissipation after completing the
discussion of the diffusion region geometry.

Many of the electrons that cross the X-point first accelerate along
the magnetosheath separatrices, producing the strong flows seen in
panel (b).  At the stagnation point their excess density %, panel (c),
leads to the converging bipolar signature in the parallel component of
the electric field $E_{\parallel}$, panel (d), that brackets the
region of positive $\mathbf{J}\boldsymbol{\cdot}\mathbf{E}$.  This
electric field ejects the electrons downstream, producing the flows
along the magnetospheric separatrices.  As part of the $J_M$ current
density that establishes the reversal in $B_L$ these same streaming
electrons have velocity components satisfying $v_M < 0$.  Hence, they
experience a $v_MB_L$ Lorentz force that pushes them in the positive
$N$ direction.  Since $B_L$ decreases when moving towards the
magnetopause, $E_N$ is eventually strong enough to counteract the
Lorentz force and again direct the electrons towards the
magnetosphere.  While continuing downstream they execute sinusoidal
oscillations (with decaying amplitude) in the $N$ direction, each
bounce accompanied by a perturbation in the net charge density and
corresponding signatures in $E_N$ and $E_{\parallel}$.  (In Figure
\ref{jdotefig} the effect is strongest on the left side of the
X-point; note the motions beginning at $L\approx -5$, $N \approx 8$ in
panel (b).  The asymmetry is likely due to the X-point's slow
diamagnetic drift \citep{swisdak03a,swisdak10a}.)  Where, due to these
undulations, $v_{N} > 0$ the electrons return energy to the fields and
produce regions with $\mathbf{J}\boldsymbol{\cdot}\mathbf{E} < 0$.
The ejection of electrons downstream from the X-point by $E_L \sim
E_{\parallel}$ contrasts sharply with the situation in symmetric
reconnection, where it is the normal magnetic field $B_N$ that rotates
the out-of-plane streaming electrons (with large $v_{M}$) into the
outflow direction.

The currents due to the electron flows also produce the large jumps
across the separatrices in the out-of-plane component of the magnetic
field $B_M$ shown in panel (e).  In symmetric reconnection $B_M$ is
quadrupolar because the electron flow is inward (towards the X-point)
along all four separatrices.  The bipolar signature in the asymmetric
case arises due to the broken symmetry \citep{karimabadi99a,tanaka08a}
and the resultant high-speed electron flows across the X-point.  Most
important for the energy conversion shown in Figure \ref{jdotefig}(a)
is the narrow (in $L$) jet of electrons with $v_N < 0$ that produces
the local reversal in $B_M$.  While this simulation includes a small
initial $B_M$, it has little effect on the system's development other
than mildly breaking the symmetry across $L=0$.  A separate simulation
with no initial $B_M$ (not shown) exhibits similar features.

To establish the dominant source of the high rate of energy
dissipation we show $S_L$, the $L$ component of the Poynting flux, in
panel (f).  Its reversal across the line $L=0$ gives the dominant
contribution to $\boldsymbol{\nabla\cdot}\mathbf{S}$ (and,
equivalently, to $\mathbf{J}\boldsymbol{\cdot}\mathbf{E}$).  Although
not shown here, plots of -$\boldsymbol{\nabla\cdot}\mathbf{S}$ and
$\mathbf{J}\boldsymbol{\cdot}\mathbf{E}$ exhibit close agreement,
supporting the claim that the system is largely in a steady-state (see
equation \ref{eqn:energy}).  Since $S_L \sim E_N B_M$, it is the
annihilation of $B_M$ between the X-point and stagnation point that
drives the large energy conversion shown in Fig.~\ref{jdotefig}(a).
Panel (g) shows the electron energy density, $\rho v_e^2/2 + 3nT_e/2$,
which peaks at the location of maximum dissipation.  The high energy
content that stretches along the magnetospheric separatrix includes a
contribution from the local electron current supporting the magnetic
field reversal as well as from the transport of energy from the
dissipation region near the X-point.

We conclude that in asymmetric reconnection the formation of a large
Hall magnetic field $B_M$ and its associated dissipation is the
dominant driver of magnetic energy release in the electron diffusion
region. This dissipation does not peak at the X-point, but rather
earthwards of it in the direction of the stagnation point.  We
emphasize that the dissipation of $B_M$ does not correspond to
reconnection of $B_M$, which would take place in the $L-M$ plane and
therefore is not accessible in the geometry of this simulation.  In
the three-dimensional simulation the region around $L=0$ in the $L-M$
plane exhibits fluctuations but there are no organized flows that
would indicate the reconnection of the $B_M$ component. This is likely
because the width of the current layer $J_N(L)$ supporting the
reversal in $B_M$ is $\lesssim 2 d_i$ in the $N$ direction and the
transit time of current-carrying electrons along the layer is of order
$\Omega_{ci}^{-1}$.

Definitively proving the existence of irreversible dissipation in
collisionless particle-in-cell simulations is not straightforward --
merely showing that regions where
$\mathbf{J}\boldsymbol{\cdot}\mathbf{E} > 0$ exist is insufficient
since reversible processes can generate such signals. The governing
Vlasov equation is, in principle, time-reversible, but it can also
lead to the development of arbitrarily complex structures in phase
space.  As the complexity increases, weaker and weaker non-ideal
processes are sufficient to cause irreversible heating and
dissipation.  Panel (h) shows the electron $v_{L}-L$ phase space for
the domain $N\in (7, 8)$, which intersects the most significant
regions of $\mathbf{J}\boldsymbol{\cdot}\mathbf{E}$.  The vertical
scale is normalized to the electron Alfv\'en speed and, as expected,
most of the plasma has been accelerated to $\approx v_{Ae0}$ within a
few $d_{e0}$ downstream of the stagnation point/X-point.  (The faint
background corresponds to hot, tenuous magnetospheric electrons.  Most
of the particles are colder, denser magnetosheath electrons that have
passed through the X- and stagnation points.)  The primary central
vortex and the secondary adjoining vortices correspond to the
oscillations in $\mathbf{J}\boldsymbol{\cdot}\mathbf{E}$. The hot,
nearly featureless beams downstream from the stagnation region suggest
that the electrons have undergone irreversible heating.  In contrast,
the ions do not undergo significant heating while traversing this
region and the analogous ion phase space (not shown) does not include
any fine-scale features.

\begin{figure}
\includegraphics[width=\columnwidth]{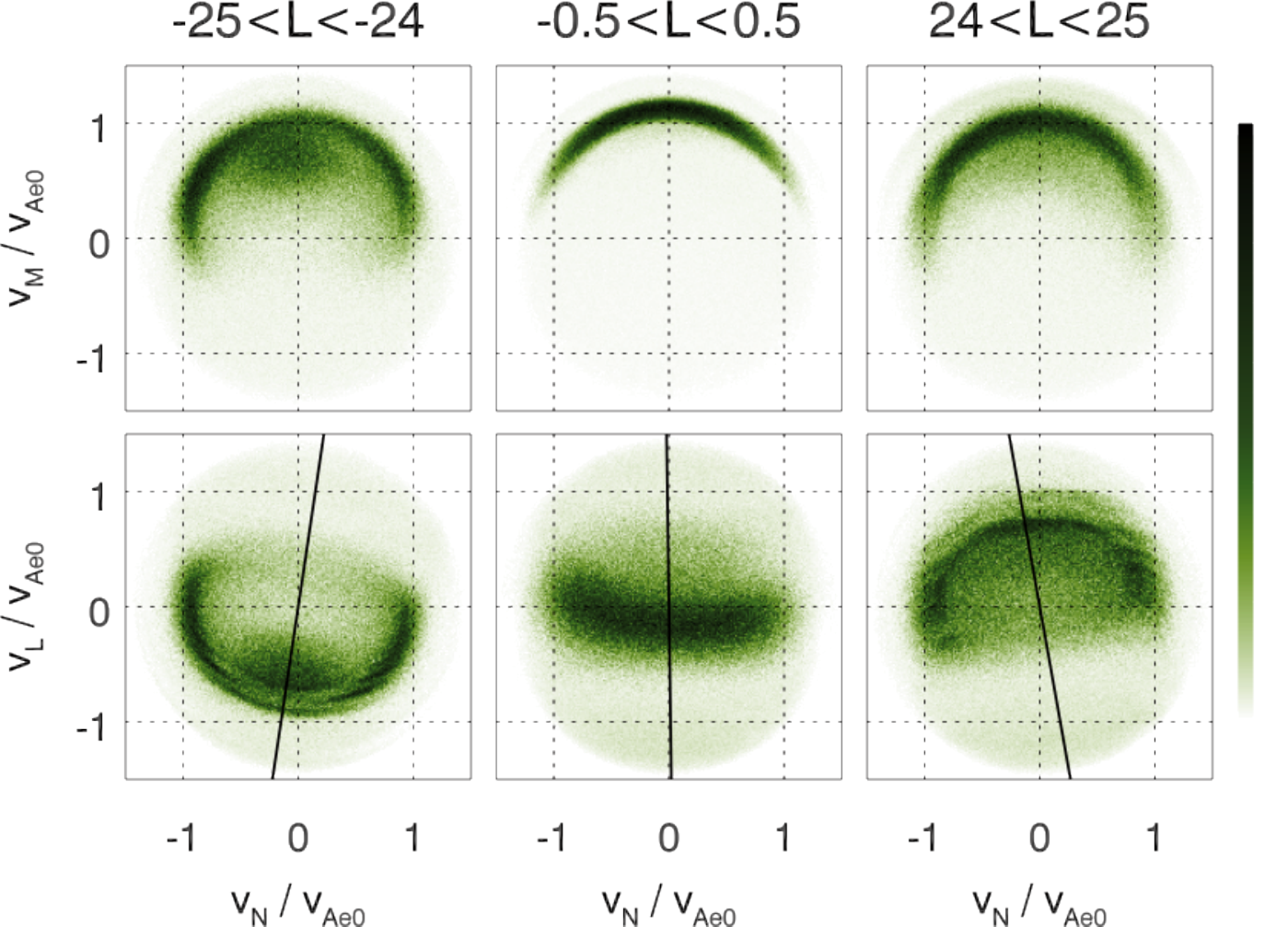}
\caption{\label{dfunc} Two-dimensional electron velocity distributions
  collected at $7<N<8$ and the far left, center, and far right of the
  domain shown in Figure \ref{jdotefig}.  The solid lines in the
  bottom panels indicate the local projection of the magnetic field
  direction.}
\end{figure}

The features of the electron diffusion region shown in Figure
\ref{jdotefig} are accompanied by signatures in the electron
velocity-space distributions.  Figure \ref{dfunc} shows
two-dimensional distributions at three locations.  Each includes
particles found in the range $7< N< 8$ while the three columns
correspond to different locations in $L$: $-25< L< -24$, $-0.5 < L <
0.5$, and $24< L< 25$.  

Cuspate motions of the electrons in the fields shown in Figure
\ref{jdotefig} produce crescent distributions in electron velocity
space.  (As in Figure \ref{jdotefig}h, hot magnetospheric electrons
comprise a faint background population.)  The top row shows the
distributions in $v_{N}-v_{M}$ space with the central panel capturing
electrons that have just entered the magnetosphere.  The crescents are
the perpendicular velocity-space features -- the local magnetic field
is nearly perpendicular to the $M-N$ plane -- predicted from
simulations \citep{hesse14a} and subsequently observed by MMS
\citep{burch16a}.  (While not strictly field-aligned, the $LMN$ axes
are good proxies for such a coordinate system since the local field
points primarily in the $L$ direction.)  They arise from magnetosheath
electrons streaming across the X-point that have their motion
deflected into the positive-$M$ direction where they form the current
density $J_M$ that supports the rotation in the reconnecting magnetic
field $B_L$.  The bottom row shows the distribution in $v_{L}-v_{N}$
space with the solid lines giving the projected orientation of the
local magnetic field.  The crescents visible in the left and right
panels, so-called parallel crescents because of their alignment with
the field, have been documented in MMS observations \citep{burch16a}
and are the result of the electrons that form the central crescents of
the top row being accelerated into the $L$ direction by the electric
field $E_L\approx E_{\parallel}$ \citep{shay16a}.

\section{Discussion}

Although our simulations necessarily include some simplifying
approximations, they agree with many features of the MMS observations
\citep{burch18a} while simultaneously providing a synoptic view.
During asymmetric reconnection with a small guide field we observe
spatially oscillatory dissipation signatures in which
$\mathbf{J}\boldsymbol{\cdot}\mathbf{E}$ changes sign over a
characteristic scale length of a few $d_{e0}$.  These features,
observed by MMS but previously unexplained, are a consequence of the
electron dynamics.  Furthermore, like MMS, we observe both
field-aligned beams along the separatrices and phase-space crescents
with varying orientations with respect to the local magnetic field,
suggesting that the essential physics of the electron flows have been
captured.

Nevertheless, discrepancies between the simulations and observations
do exist.  The oscillations are spatially stationary in the frame of
the X-point, include minimal contributions from the ions (i.e., exist
in a regime similar to electron-MHD but where kinetics play a role
\citep{shay98b}), and exhibit strongly nonlinear amplitudes.  The
latter are $\approx 3-4$ times smaller in the simulations than in the
MMS observations, likely as a consequence of the nonphysical mass
ratio.  The magnitude of $E_N$ is controlled by the ions, and hence is
insensitive to $m_i/m_e$, but $J_N\sim B_M/\delta_L$ depends on the
electron scale length $\delta_L$.  For a realistic $m_i/m_e$ the
magnitude of $\mathbf{J}\boldsymbol{\cdot}\mathbf{E}$ should increase
by $\approx \sqrt{1836/100} \approx 4$.  Also, although we argue that
the oscillations in $\mathbf{J}\boldsymbol{\cdot}\mathbf{E}$ arise
from changes in the sign of $v_{eN}$, no such variations are observed
in the MMS data.  However, $v_{eN}$ is typically much smaller than
either $v_{eM}$ or $v_{eL}$ near X-points, which make its measurement
susceptible to errors arising from small variations in the
determination of the $LMN$ coordinate system.  \cite{denton18a} have
shown that variations up to $25^{\circ}$ are possible, which is more
than sufficient to explain the discrepancy.

Three-dimensional simulations of the same event
\citep{price16a,le17a,price17a} have been previously reported.  While
the two- and three-dimensional systems generate similar large-scale
features, two factors lead us to focus on the former.  First, because
of the reduced dimensionality it is possible to track a substantially
larger number of macroparticles (in this case $10^2-10^3$ times more
per typical length scale), which significantly reduces statistical
noise. Second, while regions of intense dissipation associated with
the annihilation of the Hall magnetic field are seen in each case, the
three-dimensional simulation develops structure in the $M$ direction
associated with the lower-hybrid drift instability (LHDI) that
complicates the analysis of the dissipation mechanism but does not
appear to alter the underlying physics.

\begin{figure*}
\begin{center}
\includegraphics{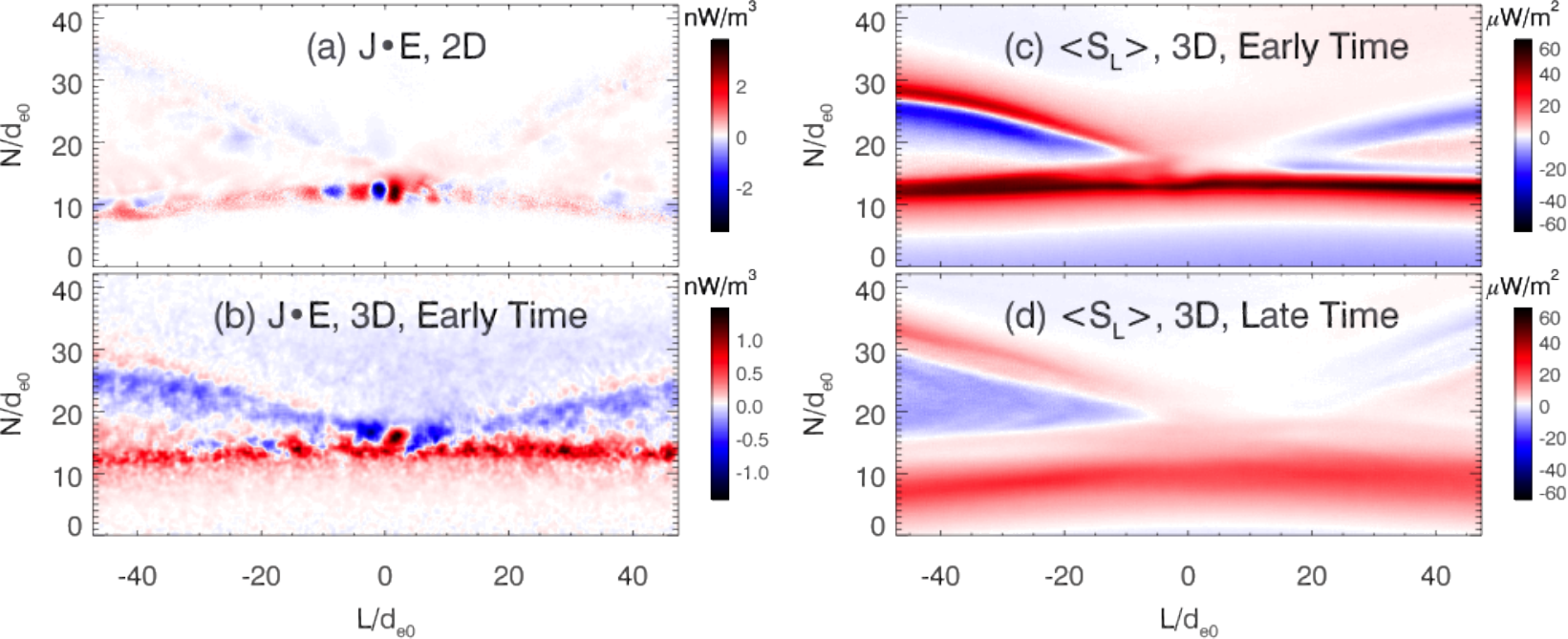}
\caption{\label{2d3d} Dissipation and Poynting flux from
  two-dimensional and three-dimensional simulations. (a)
  $\mathbf{J}\boldsymbol{\cdot}\mathbf{E}$ from the same
  high-resolution 2D simulation shown in panel a of Figure
  \ref{jdotefig}; (b) A two-dimensional slice of
  $\mathbf{J}\boldsymbol{\cdot}\mathbf{E}$ at an early time in the
  simulation, and averaged over $\approx 3d_e$ in the $M$ direction to
  reduce noise, from the lower-resolution 3D simulation presented in
  \cite{price16a}; (c) The average of $S_L$ over the entire $M$ domain
  from the same time as panel (b); (d) $\langle S_L\rangle$ from later
  in the same simulation.  The panels have been horizontally shifted
  to align the X-points.}
\end{center}
\end{figure*}

Panel (a) of Figure \ref{2d3d} shows
$\mathbf{J}\boldsymbol{\cdot}\mathbf{E}$ in the region surrounding the
X-point for the two-dimensional simulation over a slightly larger
region than that shown in panel (a) of Figure \ref{jdotefig}.  Panels
(b) and (c) show cuts from the three-dimensional simulation of
\cite{price16a}.  To reduce the random noise both panels have been
averaged over $\approx 3 d_e$ in the $M$ direction; averages over
significantly greater distances smear out significant features.  Panel
(b) comes from early in the simulation before the development of
lower-hybrid turbulence.  A comparison with panel (a) shows that both
the two- and three-dimensional simulations develop analogous features,
particularly oscillations in the sign of
$\mathbf{J}\boldsymbol{\cdot}\mathbf{E}$ aligned in the $L$ direction
and parallel to the magnetospheric separatrix.  In addition, the
three-dimensional simulation, unlike its two-dimensional counterpart,
has a significant signal along the magnetosheath separatrix. The
source of this feature is unclear.  Spacecraft on trajectories cutting
through the outflow in the $N$ direction would also observe an
oscillation in the sign of $\mathbf{J}\boldsymbol{\cdot}\mathbf{E}$.
However, except for encounters very close to the X-point the spatial
separation between the peaks would be significantly larger than the
primary signal discussed here.  Panel (c) shows the average of the
$L$-component of the Poynting vector over the entire domain in $M$,
$\langle S_L\rangle$, at the same time as shown in panel (b).
(Averages over smaller extents exhibit similar, albeit noisier,
features.)  Later in the three-dimensional simulation the lower-hybrid
drift instability drives turbulence near the X-point.  Panel (d)
displays $\langle S_L\rangle$ for this later time.  Despite the
development of the turbulence, the average structure is very similar
to that at earlier times.  In particular, the $L$-directed gradient in
$S_L$ is still present, which strongly suggests that similar physical
mechanisms occur in both the two- and three-dimensional cases.

The development of turbulence in the three-dimensional case has a
further effect.  The resulting flows in the $M-N$ plane twist $E_N$
into the $M$ direction, producing a localized $E_M \sim E_N$ that is
much larger than the reconnection electric field \cite[see Figure 10
  of][for an example of this mechanism in which the source of $E_N$ is
  the LHDI]{price16a}.  As a consequence, the intense dissipation
produced by $J_NE_N$ in the present 2D simulation will also manifest
in a comparable $J_ME_M$ term, an effect seen in both the
three-dimensional simulation and by MMS \citep{burch18a}.

Previous simulations of asymmetric reconnection with similar
parameters have observed some of the features noted here (see, for
instance, Figure 1 of \citep{pritchett09a}), but they were not fully
explained, perhaps due to excessive levels of computational noise.
The oscillatory $\mathbf{J}\boldsymbol{\cdot}\mathbf{E}$ signatures
described in this work have been seen in a similar simulation, but not
in simulations of MMS events with larger guide fields \cite[see
  Figures 2-4 of][]{cassak17a}.  This may be due to the effects,
mentioned above, of the electron diamagnetic drift.  As the strength
of the guide field increases (but $B_M < B_L$) the speed of the drift
scales with the magnitude of $B_M$.  For sufficiently strong fields
the drift can shear the vortices seen in Figure \ref{jdotefig}b and
hence may suppress the oscillatory behavior.  Exploration of this idea
requires further measurements and simulations of the electron
diffusion region in asymmetric reconnection with guide fields of
varying strength.

\begin{acknowledgments}
We gratefully acknowledge support from NASA Grants NNX14AC78G,
NNX16AG76G, and NNX16AF75.  This research uses resources of the
National Energy Research Scientific Computing Center (NERSC), a DOE
Office of Science User Facility supported by the Office of Science of
the U.S. Department of Energy.  The simulation data are archived at
NERSC and can be made available from the corresponding author by
request.
\end{acknowledgments}

%\bibliographystyle{plainnat}
%\bibliography{paper}

\end{document}